# The concentration and energetic content of floral nectar sugars: calculation, conversions, and common confusions

Jonathan G Pattrick[1,*], Jennifer Scott[1], Geraldine A Wright[1]

[1]Department of Biology, University of Oxford, John Krebs Field Station, Wytham, Oxford, OX2 8QJ

*author for correspondence: jonathan.pattrick@biology.ox.ac.uk

## Abstract

The sugar concentration of floral nectar is a key metric for describing nectar composition and a major factor influencing pollinator visitation to flowers. Across pollination biology research there are multiple approaches in use for describing nectar sugar concentration. With these different approaches there are several potential sources of confusion which, if not accounted for, can lead to errors. Further potential for error arises if researchers wish to make comparisons between the energetic content of nectars containing different ratios of sucrose, fructose and glucose. Regardless of whether concentration is measured per mole or per unit mass, the energetic content differs between the hexose sugars (glucose and fructose) and sucrose. Appropriate conversion is needed for direct comparison. Here we address these two issues with the following aims. We consolidate the literature on this topic with examples of the different methods for reporting nectar sugar concentrations, provide insight into potential sources of error, and derive equations for converting between the different ways of expressing sugar concentration for the three primary nectar sugars: sucrose, glucose and fructose. Second, we discuss the relative energetic content of sucrose, glucose, and fructose, and rationalise adjustment of 'energetic value' rather than reporting concentration directly. In this way, we hope to harmonise ongoing work in pollination ecology.

# Introduction

Floral nectar is an important resource for many animals (Nicolson 2007). The majority of the energetic value in nectar comes from the constituent sugars, which are most commonly the monosaccharides fructose and glucose, and the disaccharide, sucrose (Heinrich 1975; Nicolson 2022). Sugar composition and concentration varies considerably across plant species (Baker & Baker 1983; Abrahamczyk et al. 2017), with a mass fraction (percentage of sugars by weight) from below 5% to over 70% w/w (Nicolson & Thornburg 2007).

Sugar concentration defines the mass- or volume-specific energetic content and is consequently a key parameter for describing nectar. Across the pollination biology literature, there are a range of different methods in use for reporting sugar concentration - both for nectar and for lab-made sugar solutions for experimental testing (e.g. Whitney et al. 2008; Jones et al. 2015; Broadhead & Raguso 2021; Parkinson et al. 2022; Hemingway et al. 2024) - and there are multiple potential pitfalls associated with failing to recognise the differences between these methods. While these issues have been raised before (Bolten et al. 1979; Nicolson & Thornburg 2007), many examples can still be found in the literature of incomplete reporting of parameters and units for expressing concentration, and comparisons between studies without appropriate conversion between approaches.

Here, after reviewing the most commonly used methods for reporting nectar sugar concentration, we describe some of the problems and errors that can arise if these methods are confused. We then provide information on how to convert between these different approaches for reporting concentration. Conversion is necessary not only to facilitate accurate comparison across studies, but also is often required for calculation of the energetic content of nectar. The energetic content of nectar sugars differs depending on sugar identity, and we provide an in-depth discussion of these differences. Our manuscript builds on previous work covering these topics (Bolten et al. 1979; Kearns & Inouye 1993; Nicolson & Thornburg 2007; Prŷs-Jones & Corbet 2011) by consolidating this information in one location, providing full conversion equations for the three most common nectar sugars.

The sugar concentration of nectars is often reported as either mass of sugar as a percentage of the mass of solution (Siviter et al. 2019; Guiraud et al. 2022; Hemingway et al. 2024), mass of sugar per unit volume of solution (Broadhead & Raguso 2021; Venjakob et al. 2022), or moles of sugar per unit volume of solution (Barlow et al. 2017; Parkinson et al. 2022). In the pollination biology literature, these three approaches are commonly expressed as:

a) Percent weight per weight, (% w/w). The mass of solute (here sugar) divided by the total solution mass, expressed as a percentage. Often described as grams of solute per 100 grams of solution; however, being the quotient of two masses, % w/w is a dimensionless measure and will yield the same value regardless of the units (e.g. g, kg).

b) Percent weight per volume, (% w/v). In pollination biology this is typically the mass (in grams) of sugar per ml solution, expressed as a percentage (i.e. multiplied by 100). It is therefore equivalent to grams of sugar per 100 mL of solution.

c) Molarity (M) Defined as moles of solute per litre of solution (Lide 2004).

Two main problems associated with the use of % concentration are: 1. When reporting concentration as a percentage, not specifying the approach used (e.g. % w/w or % w/v), and 2. When comparing between studies, not adjusting for the different methods for reporting percentage concentration. Part of the confusion may arise as both % w/w and % w/v are commonly shortened to just '%' after first definition in a paper. While these give the appearance of measuring the same quantity they are two completely different measures.

For floral nectar sugars, values for % w/w and % w/v increasingly diverge as absolute concentration increases (Bolten et al. 1979, Table 1.). For instance, for fructose, a 50% w/w solution is over 61% when reported as % w/v (Table 1). Even small differences resulting from confusing the two can matter; as an example, pollinators such as bumblebees can discriminate between solutions differing by only 2% w/w (Zhou et al. 2024).

This divergence of scales has implications for calculating summary statistics of solutions of differing sugar concentrations, as the values will vary depending on how concentration is reported. For example, the mean of 10% and 70% w/w sucrose solution is 40% w/w, which is equivalent to 47.1% w/v (Table 1). However, when converted to % w/v, 10% w/w = 10.4% w/v, and 70% w/w = 94.3% w/v (Table 1). The mean of these two values is 52.4% w/v, which is not 47.1% w/v.

A further area where caution is particularly warranted is in the use of any description of sugar composition involving proportions or ratios. For example, the nectar sucrose proportion (or percentage) is a commonly-used measure to describe nectar sugar composition (e.g. Abrahamczyk et al. 2017), and is calculated as the amount of sucrose relative to the amount of total nectar sugars. Different authors have calculated this on a molar basis (Petanidou 2005) and a by mass (i.e. % w/w, % w/v or similar) basis (Nepi et al. 2010). These two approaches give very different values, and errors arise when comparing across studies without adjusting for this difference. For example, a nectar which contains 5% w/v of each of the three common sugars (i.e. 5 g sucrose / 100 ml nectar, 5 g fructose / 100 ml nectar and 5 g glucose / 100 ml nectar) has a nectar sucrose percentage, calculated by mass, of 33.3%, ($\frac{5}{5+5+5} \times 100$). In contrast, the nectar sucrose percentage calculated on a molar basis is 20.8%, ($\frac{5/342.3}{(5/342.3) + ((5+5)/180.155)} \times 100$), where 342.3 and 180.155 are the molecular weights of sucrose and the two hexoses respectively.

**Table 1.** Comparison between different methods for reporting the sugar concentration of nectar. For solutions comprising a single sugar (sucrose (Suc), glucose (Gluc) or fructose (Fruc)) ranging in concentration from 0 to 70% w/w, the respective concentration is given in % w/v (g / 100 ml solution) and molarity (M, moles / litre solution). Conversions were calculated using equations 1, 2 and 3.

| Concentration (% w/w) | Concentration (% w/v)* (g solute per 100 ml solution) | | | Concentration (M) | | |
|---|---|---|---|---|---|---|
| | Suc | Gluc | Fruc | Suc | Gluc | Fruc |
| 0 | 0 | 0 | 0 | 0 | 0 | 0 |
| 10 | 10.4 | 10.4 | 10.4 | 0.303 | 0.576 | 0.576 |
| 20 | 21.6 | 21.6 | 21.6 | 0.631 | 1.20 | 1.20 |
| 30 | 33.8 | 33.7 | 33.8 | 0.988 | 1.87 | 1.88 |
| 40 | 47.1 | 46.9 | 47.1 | 1.37 | 2.60 | 2.61 |
| 50 | 61.5 | 61.2 | 61.5 | 1.80 | 3.40 | 3.41 |
| 60 | 77.2 | 76.7 | 77.1 | 2.26 | 4.26 | 4.28 |
| 70 | 94.3 | 93.5 | 94.0 | 2.75 | 5.19 | 5.22 |

*The nature of the difference between % w/w and % w/v can be illustrated using the following example. A 50% w/w fructose solution can be made up by dissolving 50 g of fructose in 50 g of water. As 50 g of water is (at room temperature and pressure) approximately 50 ml, the addition of fructose to this will raise the volume of the resulting solution to 81.35 ml. Given we have 50 g in 81.35 ml, when considering the concentration as % w/v (i.e. g / 100 ml solution), this will be greater than 50% w/v, giving the value of $\frac{50}{81.35} \times 100$ = 61.5% w/v. Consequently, sugar solutions can have a concentration >100% when reported as % w/v, but this is not possible for % w/w.

There are several other ways of expressing concentration that are less widely used in pollination ecology. Bolten et al. (1979) give a second version of % w/v which they define as grams of solute per 100 ml *solvent*. Although Bolten et al. (1979) do provide a reference for a study purportedly using these units, this appears to have been cited in error as the study in question states that "sugar solutions were made up on a weight sugar per volume *solution* basis", i.e. grams solute per 100 ml *solution* (Stiles 1976). However, we have found at least one paper which, based on the sucrose concentrations reported, does appear to use this measure (Grassl et al. 2018). While use of % w/v (g solute per 100 ml solution) seems to be more common, this is difficult to assess as typically the approach used is not specified (see Supplementary Information). Therefore we suggest where researchers do use % w/v (and indeed % w/w) it would be prudent to state exactly what is meant by these measures and specify the units used (e.g. Bertazzini & Forlani 2016; Brown & Brown 2020; Rering et al. 2020), to avoid any potential confusion. A nectar of 40% w/v sugar concentration is ambiguous, but 40 g sugar / 100 ml solution is not. Unless directly specified, throughout the rest of this manuscript, any mention of % w/v is equivalent to grams solute per 100 ml solution, and % w/w is defined as above.

Occasionally, studies report concentration of sugar solutions in % v/v. This refers to volume of the compound (here the dry solute: sugar) in ml in 100 mL of the final solution. While % v/v is an appropriate measure when mixing two liquids, we recommend that, where possible, these units are avoided for reporting sugar concentration. This is because of the difficulty in accurately measuring volume of a dry solute. An additional problem is that the density and therefore volume of the solute (sugar) depends on the crystal size and how the sugar is handled (compare loose density versus tapped density, where the sugar is subjected to vibration) during measurement (Santos et al. 2018). Unless the density of the sugar crystals is given (which we have not seen in any paper in the field of pollination biology using % v/v), it is not possible to accurately convert between % v/v and other parameters.

Nectar sugar concentration is often measured using refractometers. These commonly give concentration either directly in % w/w sucrose equivalents, or on the Brix scale which, for sucrose, is equivalent (Bolten et al. 1979). Fructose- and glucose-specific refractometers do exist as the refractive indices for the hexoses (glucose and fructose) are very slightly lower than those for sucrose (Fucaloro et al. 2007). At low concentrations this difference is negligible, but it increases slightly with increasing concentration (% w/w). For most purposes in pollination biology the difference is only going to have a small impact. For example, using a Brix scale (sucrose) digital refractometer (Atago PAL-1) zeroed with RO water, we measured the concentration of a 20% w/w fructose (Sigma) solution as 19.7% and a 40% w/w fructose solutions as 39.1%. For comparison, a 40% w/w sucrose solution measured as 40.1%. A more significant source of error associated with measuring nectar in the field is the potential presence of other non-sugar nectar constituents such as amino acids and inorganic ions. While typically present at low levels in nectar (Nicolson & Thornburg 2007), concentrations occasionally can get much higher, for example 80 mM total amino acids (Göttlinger & Lohaus 2024) and circa. 0.33 M for potassium (Waller et al. 1972). These can alter the refractive index by as much 3% w/w equivalents when present in high concentrations (Inouye et al. 1980; Hiebert & Calder 1983). Additionally, the refractive index of a solution is also affected by temperature, so readings should ideally be taken at/near to the calibration temperature for the refractometer, or corrected as necessary (e.g. Roubik et al. 1995).

Nectar sugars are important to pollinators as they are used as an energy source. In experiments where researchers are interested to know how the value of nectar influences foraging behaviour, it can be particularly insightful to convert concentration into energetic value. If the volumetric energetic content (e.g. joules per μl) is required, then at this point it is necessary to convert any concentrations reported in % w/w to % w/v (or molarity). Here, the choice of parameters is crucial as the energetic content of sucrose differs from that of the hexose sugars (glucose and fructose) both when concentration is measured on a by weight basis (% w/w and % w/v) and by molarity (Fleming et al. 2004; Nicolson & Thornburg 2007; Brown et al. 2008). Although the energetic content of sucrose is widely available from the pollination biology literature, there is some variability depending on the source (Heinrich 1975; Harder 1986; Balfour et al. 2015). We would therefore recommend caution when quoting the energetic content of sucrose unless the source includes a specific explanation of how this is derived. In several cases a single incorrect value is cited multiple times by subsequent works. In addition, finding values for fructose and glucose is more difficult (but see Brown et al. 2008; Schmid et al. 2011; Pille Arnold et al. 2024).

In the following sections, we present formulae for converting between different approaches for reporting nectar sugar concentration, derive the energetic content of sucrose, fructose and glucose, and rationalise adjustment of nectar concentration based on the 'energetic value'.

# Converting between approaches for reporting sugar concentration

Both % w/v and molarity are measures of dry solute made up to a known final solution volume. The conversion requires converting between moles and mass (for which the molecular weight is needed) and then adjusting for the volume difference between the two measures (100 ml for % w/v and 1 L for molarity).

1. Converting between % w/v (g / 100 ml solution) and molarity:

$$C_M = 10\left(\frac{C_v}{Mr}\right), \qquad (1)$$

where $C_M$ is the molar concentration in moles per litre, $C_v$ is the concentration in % w/v (g / 100 ml solution) and $Mr$ is the molecular weight of the respective sugar (342.3 for sucrose and 180.155 for fructose and glucose). Given that the molecular weight of sucrose is a little under double that of fructose and glucose, when converted to % w/v, a 1 M solution of sucrose will be a little under double the % concentration of a 1 M glucose or fructose solution (34.2% w/v for sucrose versus 18.0% w/v for fructose and glucose).

Converting between either molarity or % w/v to % w/w requires the concentration-specific density of the respective solution. To predict density from concentration, and thereby provide equations for converting between % w/w and % w/v, we used the lm function in R Version 4.3.2 (R Core Team 2023) to fit second-order polynomials to the density data for sucrose, fructose and glucose solutions at 20°C given in the CRC handbook of Chemistry and Physics (Lide 2004). To convert from concentration in % w/w to % w/v ($C_v$) we can use $C_v = C_w \rho_w$, where, $\rho_w$ is the density of the solution at concentration $C_w$ in % w/w and can be predicted using $\rho_w = a_1 + a_2 C_w + a_3 C_w^2$, where the parameters $a_1$, $a_2$, and $a_3$ are dependent on the sugar being used and are given in Table 2. Combined, this gives equation 2:

2. Converting to % w/v (g / 100 ml solution) from % w/w (single sugar solutions)

$$C_v = C_w(a_1 + a_2 C_w + a_3 C_w^2) \qquad (2)$$

**Table 2.** Parameters for calculating the density of sugar solutions of a given concentration in % w/w for converting to % w/v (g solute / 100 ml solution) from % w/w (Equation 2).

| Sugar | Parameters | | |
|---|---|---|---|
| | $a_1$ | $a_2$ | $a_3$ |
| Sucrose | 0.998709 | 3.7430E-03 | 1.7639E-05 |
| Glucose | 0.998442 | 3.7390E-03 | 1.5410E-05 |
| Fructose | 0.998371 | 3.8550E-03 | 1.5230E-05 |

*The units for $a_1$, $a_2$ and $a_3$ are g / 100 ml

After calculating the density and converting to % w/v using equations 2 and 3, we fitted similar second-order polynomials to obtain formulae for the density of sucrose, glucose and fructose solutions where the concentration is given in % w/v to allow for transformation back to % w/w. Given the density of the solution, the concentration in % w/w is $C_w = C_v / \rho_v$, where $C_v$ is the concentration in % w/v and $\rho_v$ is the density of the solution at concentration $C_v$ (in % w/v). The respective solution densities can be predicted using: $\rho_v = b_1 + b_2 C_v + b_3 C_v^2$ where the parameters $b_1$, $b_2$, and $b_3$ are again sugar-dependent and are given in Table 3. Combined, this gives equation 3:

3. Converting to % w/w from % w/v (g / 100 ml solution) (single sugar solutions)

$$C_w = \frac{C_v}{b_1 + b_2 C_v + b_3 C_v^2} \qquad (3)$$

**Table 3.** Parameters for calculating the density of sugar solutions of a given concentration in % w/v for converting to % w/w from % w/v (g solute / 100 ml solution) (Equation 3).

| Sugar | Parameters | | |
|---|---|---|---|
| | $b_1$ | $b_2$ | $b_3$ |
| Sucrose | 0.998138 | 3.8765E-03 | -1.8716E-06 |
| Glucose | 0.998226 | 3.8084E-03 | -1.9924E-06 |
| Fructose | 0.998253 | 3.9021E-03 | -2.2878E-06 |

* The units for $b_1$ are g / 100 ml, $b_2$ is dimensionless, and for $b_3$ 100 ml / g.

All fitted polynomial models had adjusted R² > 0.999.  When compared to the original values from the CRC handbook (Lide 2004) the errors for the predicted density for a given sugar concentration in % w/w were all lower than 0.09% and agree with other published values (Bates 1942).  These conversions are valid for a temperature of 20°C, though any difference in density from working at a higher or lower temperature (e.g. between 10°C to 40°C) is sufficiently small (Darros-Barbosa et al. 2003) that these conversions will be accurate for most situations in pollination biology.  We provide an R script/function for converting between the different parameters for reporting concentration as Supplementary Material.

Equations 1 through 3 are only valid for single-sugar solutions (i.e. a binary solution of one sugar in water). Nectar typically contains glucose, fructose and sucrose in varying proportions (as well as occasionally some minor sugars as well). For trinary or quaternary solutions (i.e. multi-sugar solutions with two or three sugars respectively) converting between molarity and % w/v (g / 100 ml solution) requires equation 1 (above) to be applied to each sugar separately.  For instance, a nectar which is 0.5 M sucrose, 1 M glucose and 1 M fructose will contain $\frac{0.5}{10} \times 342.3 = 17.12\%$ w/v sucrose, $\frac{1}{10} \times 180.155 = 18.02\%$ w/v glucose and $\frac{1}{10} \times 180.155 = 18.02\%$ w/v fructose.

Converting between % w/w and % w/v for multi-sugar solutions is more complicated as it requires the density of the combined solution. If this cannot be directly measured, then we suggest using the density of a sucrose solution equivalent to the sum of the concentrations of the individual sugars. Our work making a range of sugar solutions suggests this this gives a reasonable approximation to the true density.

To convert to % w/v from % w/w for each individual sugar in a multi-sugar solution we can use Equation 4, applied to each sugar separately.

4. Converting to % w/v (g / 100 ml solution) from % w/w (multi-sugar solutions)

$$C_{v\ sugar} = C_{w\ sugar}(a_1 + a_2 C_{w\ combined} + a_3 C^2_{w\ combined}) \quad (4),$$

where $C_{w\ combined}$ is the sum of the concentration of the individual sugars in % w/w, $C_{w\ sugar}$ is the concentration of the sugar in question in % w/w, $C_{v\ sugar}$ is the concentration of the sugar in question in % w/v (g / 100 ml solution), and the parameters $a_1$, $a_2$, and $a_3$ are the values for sucrose from Table 2. As with Equation 2, the total solution density is given by the part of Equation 4 in parenthesises.

To convert to % w/w from % w/v for each sugar in a multi-sugar solution we can use equation 5, again applied to each sugar separately.

5. Converting to % w/w from % w/v (g / 100 ml solution) (multi-sugar solutions)

$$C_{w\,sugar} = \frac{C_{v\,sugar}}{b_1 + b_2 C_{v\,combined} + b_3 C_{v\,combined}^2} \quad (5),$$

where $C_{v\,combined}$ is the sum of the concentration of the individual sugars in % w/v (g / 100 ml solution), $C_{w\,sugar}$ and $C_{v\,sugar}$ are as in equation 4, and the parameters $b_1$, $b_2$, and $b_3$ are the values for sucrose from Table 3. As with Equation 3, the density is given by the denominator of Equation 5. Some caution is needed when converting these multi-sugar solutions between % w/w and % w/v as values vary depending on the concentration relative to the other sugars present. We present a full worked example for conversion between % w/w and % w/v in the supplementary information.

## The energetic content of nectar sugars

We calculated the energetic content of sucrose, glucose and fructose (Table 4) as the heat of combustion at 25°C and 100 kPa (i.e. atmospheric pressure) using Hess's law and the enthalpy of formation for the three sugars and their combustion products $CO_2$ and $H_2O$ (see Supplementary Information). Enthalpies of formation were obtained from the CRC Handbook of Chemistry and Physics (Lide 2004).

**Table 4.** The energetic content of sucrose, glucose and fructose, expressed in four different units.

| Sugar | Energy content (kJ / mole) | Energy content (kcal / mole) | Energy content (J / mg) | Energy content (cal / mg) |
|---|---|---|---|---|
| Sucrose | 5640.15 | 1348.0 | 16.48 | 3.94 |
| Glucose | 2802.74 | 669.87 | 15.56 | 3.72 |
| Fructose | 2810.44 | 671.71 | 15.60 | 3.73 |

To illustrate how energetic content of sugar solutions varies with sugar identity and, how this difference also depends on the method for reporting sugar concentration, we also calculated the energetic content for a standard (nectar-relevant) volume of 1 μL of equal concentrations of sucrose, glucose and fructose solutions for several measurement scales (Table 5). When solution concentration is measured in molarity, the energetic content of sucrose solution is approximately double that of an equimolar glucose or fructose solution. This is because sucrose is a disaccharide, formed from two monosaccharide subunits: one molecule each of glucose and fructose. However, when measured by weight (% w/v or % w/w), fructose and glucose have around 95% the energetic content of a sucrose solution of the same concentration (Table 5). This difference can be explained by the fact that the molecular weight of sucrose is a little under double that of glucose or fructose; the formation of sucrose from fructose and glucose results in the loss of a water molecule. The sum of the molecular weight of one fructose and one glucose molecule (180.155 for each) is therefore equivalent to that of the molecular weight of water (18.015) + the molecular weight of sucrose (342.3): (180.155 x 2) ≈ (342.3 + 18.015). Given two hexose molecules weigh more than, but have approximately equivalent energetic content to, one sucrose molecule, two hexose molecules have a lower relative energetic content than one sucrose molecule: $\frac{\frac{1}{(180.155 \times 2)}}{\frac{1}{342.3}} \times 100 = 95.0\%$. The actual value is slightly lower than 95% (Table 5) as the energetic content of sucrose is very slightly over double that of either glucose or fructose. The small differences in proportional energetic content between % w/v and % w/w (Table 5) are owing to the differences in solution densities between the three sugars. Although here we chose a specific concentration for each approach for illustrative purposes, the relative energetic content of the three sugars is independent of concentration for molarity and % w/v, and changes only very slightly with concentration for % w/w.

Often, owing to experimental limitations, the relative proportions of nectar sugars are not recorded when measuring nectar composition, just the percentage mass of sugars (Bailes et al. 2018; Symington & Glover 2024). Therefore, it is worth bearing in mind that, depending on the exact ratios of the different nectar sugars, nectars with the same concentration (% w/w) may differ in energetic content by up to approximately 5%. Indeed, it is not uncommon to find sucrose- or hexose-dominated nectars (Abrahamczyk et al. 2017).

A null hypothesis of interest to pollination biologists may involve comparing equicaloric nectars, i.e. those with equivalent energetic content (Brown et al. 2008). For this purpose, we suggest using an adjusted sucrose equivalents molarity scale as proposed by Nicolson and co-workers (Fleming et al. 2004, 2008; Leseigneur & Nicolson 2009). On this adjusted scale, the molar concentration of fructose and glucose is halved so that, for instance, a 2 M glucose solution = 1 M sucrose equivalents. The three nectar sugars have almost identical energetic content on this scale (Table 5). Without this adjustment, any comparison between the hexoses and sucrose will be biased by energetic differences between these sugars.

**Table 5.** Comparative energetic values for 1 μL of sucrose, glucose and fructose at equal concentrations on four different measurement scales.

| Scale | Concentration | Energetic content (J / μl) | | | Relative % energetic content of hexoses compared to sucrose |
|---|---|---|---|---|---|
| | | Sucrose | Glucose | Fructose | |
| Molarity | 1 M | 5.64 | 2.80 | 2.81 | 49.7% (Gluc) 49.8% (Fruc) |
| % w/v (g solute / 100 ml solution) | 30% | 4.94 | 4.67 | 4.68 | 94.4% (Gluc) 94.7% (Fruc) |
| % w/w (g solute / 100 g solution)* | 30% | 5.57 | 5.25 | 5.28 | 94.2% (Gluc) 94.7% (Fruc) |
| Sucrose equivalent molarity | 1 M | 5.64 | 5.61 | 5.62 | 99.4% (Gluc) 99.7% (Fruc) |

Values were rounded to 3 significant figures after calculation. *N.B. For molarity and % w/v (g solute / 100 ml solution) and for a given sugar, a doubling of the sugar concentration will double the volumetric energetic content; however, this is not the case for % w/w, as % w/w describes concentration relative to total solution mass, not volume: e.g. 1 μL of 60% w/w sucrose solution *will not* have double the energetic content of 1 μL of 30% w/w sucrose solution. Consequently, the values for 30% w/w in this table *should not* be used to interpolate the volumetric energetic content of solutions at other concentrations in % w/w.

In some situations, such as investigating gustatory acuity, the number of molecules of a compound interacting with taste sensilla is of importance, and here using equimolar solutions may be the most appropriate option (Miriyala et al. 2018; Parkinson et al. 2022). Furthermore, pollinators may also use sugars for purposes other than energy generation, such as a substrate for wax production in bees (Hepburn et al. 2014).

It should be noted that these energetic values are all equivalent to the enthalpy of combustion for each sugar. This is typically the value used in pollination biology (e.g. Balfour et al. 2015); however, from the perspective of the pollinator, this makes the unrealistic assumption that energy conversion through respiration is 100% efficient. An alternative approach is to calculate the number of molecules of ATP that can be produced by each sugar. However, ATP generation will likely vary on several species-specific factors – for instance the extent to which respiration is aerobic or anaerobic (Harrison & Roberts 2000). Finally, not all flower visitors can metabolise all sugars (Allsopp et al. 1998; Fleming et al. 2008; Parkinson et al. 2022).

## Summary

Here, we have presented methods for converting between different ways of measuring concentration, illustrated the derivation of the energetic content of different nectar sugars and how energetic content depends on the method used for reporting concentration. For clarity and avoiding potential errors, we suggest the following steps when working with nectar sugar concentrations:

1. If reporting sugar concentrations as a percentage, always specify the method used (e.g. % w/w, % w/v) and, for % w/v, specify the units (e.g. g / 100 ml solution) as well.
2. When comparing nectar concentrations or sugar proportions between studies, ensure the values/measures are on equivalent scales, and transform as necessary. This is particularly the case when using any value expressed as a percentage, proportion, or ratio.
3. In situations where the energetic value of nectar is of importance, comparisons between nectars should be made on an equicaloric basis, not equi-molar or mass. Sucrose-equivalent molarity is well-suited for this purpose.

In summary, we hope that by consolidating information from several sources and highlighting the potential issues arising from the variety of methods and units in use for measuring nectar sugar concentration, this manuscript will be a useful resource for established researchers and a guide for those starting out in pollination biology.

## Acknowledgements

We gratefully acknowledge the funding bodies supporting our work: The Leverhulme Trust Grant no; RPG-2020-393, The Natural Environment Research Council grant NE/V012282/1, and the Biotechnology and Biological Sciences Research Council grant BB/S000402/1. We also thank David Labonte, Hamish Symington, Joseph Gillson, Angharad Thomas-De Paul and two anonymous reviewers for helpful comments on earlier versions of this manuscript

## Data availability

All data and code supporting this manuscript will be made available as supplementary files when published. In the meantime, please contact the primary author if you would like the files.

## Conflicts of Interest

The authors declare there are no conflicts of interest.

Supplementary Information for: The concentration and energetic content of floral nectar sugars: calculation, conversions, and common confusions

Jonathan G Pattrick[1,*], Jennifer Scott[1], Geraldine A Wright[1]

[1]Department of Biology, University of Oxford, John Krebs Field Station, Wytham, Oxford, OX2 8QJ

1. Calculating the energetic content of sucrose, glucose, and fructose.

The most common values given for the energetic content of nectar sugars are those derived from the enthalpy of combustion ($\Delta_cH°$) for each of these compounds (Brown et al. 2008; Schmid et al. 2011; Balfour et al. 2015). The enthalpy of combustion can be calculated from the enthalpy of formation using Hess's law and the known enthalpy of formation ($\Delta_fH°$) for the combustion products ($CO_2$ and $H_2O$). These are directly available from the CRC Handbook of Chemistry and Physics (Lide 2004) with $\Delta_fH°$ for sucrose = -2226.1, glucose = -1273.3, fructose = -1265.6, $CO_2$ = -393.510 and $H_2O$ = -285.830 kJ/mol, all at a temperature of 298.15 K (Lide 2004).

Hess's law states that $\Delta_cH° = \sum(\Delta_fH°$ of the products$) - \sum(\Delta_fH°$ of the reactants$)$ (Lide 2004).

For fructose and glucose the combustion reaction is:

$C_6H_{12}O_6 + 6O_2 \rightarrow 6CO_2 + 6H_2O$ 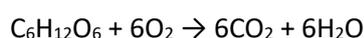

And for sucrose the combustion reaction is:

$C_{12}H_{22}O_{11} + 12O_2 \rightarrow 12CO_2 + 11H_2O$ 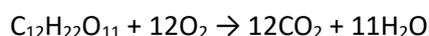

Hence for the three sugars the enthalpy of combustion can be calculated as follows:

$\Delta_cH°$ (sucrose) = [12(−393.51) + 11(−285.83)]−(−2226.1) = -5640.15 kJ / mol

$\Delta_cH°$ (glucose) = [6(−393.51) + 6(−285.83)]−(−1273.3) = -2802.74 kJ / mol

$\Delta_cH°$ (β-D-fructose) = [6(−393.51) + 6(−285.83)]−(−1265.6) = -2810.44 kJ / mol

Energy in kJ / mol can be converted to Calories (kcal) as there are 4.184 kJ per kcal (using the definition for the thermochemical calorie) (Lide 2004).

## 2. Converting between % w/w and % w/v (g / 100 ml solution) for solutions with multiple sugars

Here we give worked examples for converting between % w/w and % w/v (g / 100 ml solution) for sugar solutions containing multiple sugars, using equations 4 and 5 from the main text.

As an example for converting to % w/v from % w/w, consider a nectar which contains 10% w/w sucrose, 15% w/w glucose and 12% w/w fructose. The total sugar content (in g) in 100 g of solution is equivalent to $C_{w\ combined} = 10 + 15 + 12 = 37$ and, using Equation 4:

for sucrose, $C_{w\ sugar} = 10$, therefore $C_{v\ sugar} = 10 \times (0.998709 + 3.7430\text{E}^{-3} \times 37 + 1.7639\text{E}^{-5} \times 37^2) = 11.6\%$ w/v,

for glucose, $C_{w\ sugar} = 15$, therefore $C_{v\ sugar} = 15 \times (0.998709 + 3.7430\text{E}^{-3} \times 37 + 1.7639\text{E}^{-5} \times 37^2) = 17.4\%$ w/v.

For fructose, $C_{w\ sugar} = 12$, therefore $C_{v\ sugar} = 12 \times (0.998709 + 3.7430\text{E}^{-3} \times 37 + 1.7639\text{E}^{-5} \times 37^2) = 13.9\%$ w/v.

Here, as % w/v is equivalent to g / 100 ml (and therefore mg / 100 µl) of solution, these values can then be used in combination with Table 4 (i.e. 16.48, 15.56 and 15.60 J mg$^{-1}$ for sucrose, glucose and fructose respectively) to work out the energetic content per µl of this nectar.

$$\frac{16.48 \times 11.6 + 15.56 \times 17.4 + 15.60 \times 13.9}{100} = 6.80 \text{ J / µl}$$

As an example for converting to % w/w from % w/v, consider a nectar which contains 20% w/v sucrose, 14% w/v glucose and 18% w/v fructose. The total sugar content (in g) in 100 ml of solution is equivalent to $C_{v\ combined} = 20 + 14 + 18 = 52$ and, using Equation 5:

for sucrose, $C_{v\ sugar} = 20$, therefore, $C_{w\ sugar} = \frac{20}{0.998138 + 3.8765\text{E}^{-3} \times 52 - 1.8716\text{E}^{-6} \times 52^2} = 16.7\%$ w/w,

for glucose, $C_{v\ sugar} = 14$, therefore, $C_{w\ sugar} = \frac{14}{0.998138 + 3.8765\text{E}^{-3} \times 52 - 1.8716\text{E}^{-6} \times 52^2} = 11.7\%$ w/w,

for fructose, $C_{v\ sugar} = 18$, therefore, $C_{w\ sugar} = \frac{18}{0.998138 + 3.8765\text{E}^{-3} \times 52 - 1.8716\text{E}^{-6} \times 52^2} = 15.1\%$ w/w.

As noted in the main text, the values calculated for each individual sugar in these conversions will vary depending on the concentration of the other sugars. For example, given a nectar which is 20% w/w sucrose, 20% w/w glucose, and 20% w/w fructose, the sucrose concentration of this nectar when converted to % w/v will be 25.74% w/v. However, for a similar nectar but with no fructose in (i.e. 20% w/w sucrose, 20% w/w glucose and 0% w/w fructose), the sucrose concentration will be 23.53% w/v.

## 3. Methods for reporting sugar concentration in % w/v

Concentration reported as % w/v can refer to grams of solute in 100 ml of *solution* or, grams of solute in 100 ml *solvent* (Bolten et al. 1979). For clarity, a 20% w/v (g solute / 100 ml solution) sucrose solution would be made by diluting 20 g of sucrose in water, made up to a final solution volume of 100 ml. Meanwhile, a 20% w/v (g solute / 100 ml solvent) sucrose solution would be made by adding 20 g of sucrose to 100 ml of water. For this second approach the final solution volume would therefore be >100 ml.

For very dilute solutions (e.g. < 0.1%) the difference between the two approaches is minimal. However, pollination ecology often concerns high solute concentrations, where the sugar forms a significant proportion of the total volume of the solution.

Although % w/v (g solute / 100 ml solution) would seem to be the more commonly used measure (Lide 2004); we assessed this with a small literature search. Using Google Scholar we used the search terms "%, w/v, nectar", and then "%, w/v, nectar, pollinator" selecting the first 50 peer-reviewed papers (25 from each search, but avoiding replications) published from 2014 – 2024. We then classified the papers based on the method used for reporting concentration. Sometimes the method wasn't explicitly stated, but could be inferred.

Out of the 50 papers, we classified 10 as using % w/v (g solute / 100 ml solution), whereas one used % w/v (g solute / 100 ml solvent). We were unable to distinguish which of the two methods was used for the remaining 39 papers. This suggests that % w/v (g solute / 100 ml solution) is indeed more common, but not the only approach used. We therefore we recommend explicitly stating the approach used.

For completeness, we also include equations to convert from % w/v (g sugar / 100 ml *solvent*) to the other measures considered. Concentration in % w/v (g sugar / 100 ml solvent) can be converted to % w/w using SI Equation 1:

Converting to % w/w from % w/v (g sugar / 100 ml solvent)

$$C_w = \frac{C_{vsolvent}}{100 + C_{vsolvent}} \times 100, \qquad \text{SI Equation 1}$$

where $C_{vsolvent}$ is the concentration in % w/v (g sugar / 100 ml solvent) and $C_w$ is the concentration in % w/w, as in the main text. This can be converted to % w/v (g sugar / 100 ml *solution*) or other measures using the equations in the main text. The reverse conversion is:

Converting to % w/v (g sugar / 100 ml solvent) from % w/w

$$C_{vsolvent} = \frac{C_w}{100 - C_w} \times 100. \qquad \text{SI Equation 2.}$$

To illustrate the difference in values given by these methods for reporting concentration, we present values for sucrose concentration for % w/v (g sucrose / 100 ml solution), % w/w and % w/v (g sucrose / 100 ml solvent) (SI Table 1).

**SI Table 1.** A comparison between the concentrations of sucrose solutions given as % w/v (g solute / 100 ml solution), % w/w (g solute / 100 g solution) and % w/v (g solute / 100 ml solvent).

| Concentration (% w/v) (g sucrose / 100 ml solution) | Concentration (% w/w) (g sucrose / 100 g solution) | Concentration (% w/v) (g sucrose / 100 ml solvent) |
|---|---|---|
| 0 | 0 | 0 |
| 10 | 9.6 | 10.7 |
| 20 | 18.6 | 22.9 |
| 30 | 27.0 | 36.9 |
| 40 | 34.8 | 53.3 |
| 50 | 42.1 | 72.8 |
| 60 | 49.0 | 96.2 |
| 70 | 55.5 | 124.9 |
| 80 | 61.7 | 161.2 |
| 90 | 67.6 | 208.4 |